  \documentclass[10pt,oneside,a4paper]{article}

  \usepackage[
  backend=biber,
  style=bwl-FU,
  sorting=nyt
  ]{biblatex}
  \usepackage[pdftex]{graphicx}
  \graphicspath{{../pdf/}{../res/}}
  \DeclareGraphicsExtensions{.pdf,.jpeg,.png}
  \usepackage{amsmath}
  \usepackage{amssymb}
  \usepackage{array}

  \usepackage{pgfplots}
  \usepackage{pgfplotstable}
  \usepgfplotslibrary{fillbetween} 
  \usepackage{tikz}
  \usepackage{caption}
  \usepackage{subcaption}

  \pgfplotsset{width=7cm,compat=1.3}
  \usepackage{csvsimple}
  \usepackage{multirow}
  \usepackage{dcolumn}

  \usepackage{booktabs, makecell}
  \setcellgapes{3pt}
  \usepackage{siunitx} 

  \usepackage{url}

  \usepackage{authblk}
  \usepackage[title]{appendix}%
  \usepackage{comment}

  \usepackage{amsthm} 
  
  \newtheorem{definition}{Definition}

  \usepackage{titlesec} 
  \titleformat{\section}
  {\normalfont\Large\bfseries}{\thesection}{1em}{}

\begin{document}

  \title{Bimodal Dynamics of the Artificial Limit Order Book Stock Exchange with Autonomous Traders}
  
  \author[1]{Matej Steinbacher\thanks{matej.steinbacher@gmail.com}}
  \author[2]{Mitja Steinbacher\thanks{mitja.steinbacher@kat-inst.si}}
  \author[3]{Matjaž Steinbacher\thanks{corresponding author}}
  
  \affil[1]{%
  Independent Researcher}
  \affil[2]{%
  Faculty of Law and Business Studies, Catholic Institute, Ljubljana, Slovenia}
  \affil[3]{%
  Fund for Financing the Decommissioning of the Krško Nuclear Power Plant and Disposal of Radioactive Waste, Krško, Slovenia}


  %
  %

  \maketitle

  \begin{abstract}
    This paper explores the bifurcative dynamics of an artificial stock market exchange (ASME) with endogenous, myopic traders interacting through a limit order book (LOB). We showed that agent-based price dynamics possess intrinsic bistability, which is not a result of randomness but an emergent property of micro-level trading rules, where even identical initial conditions lead to qualitatively different long-run price equilibria: a deterministic zero-price state and a persistent positive-price equilibrium. The study also identifies a metastable region with elevated volatility between the basins of attraction and reveals distinct transient behaviors for trajectories converging to these equilibria. Furthermore, we observe that the system is neither entirely regular nor fully chaotic. By highlighting the emergence of divergent market outcomes from uniform beginnings, this work contributes a novel perspective on the inherent path dependence and complex dynamics of artificial stock markets.
  \end{abstract}


  \section{Introduction}

  It was shown in \cite{steinbacher2025mathematical} that the directionality and magnitude of price dynamics of an artificial stock market exchange (ASME) with endogenous, myopic traders, fixed total money supply and the fixed total number of stocks that are equally spread among traders is determined by the aggregate money supply, which defines the direction of the equilibrium price, and by market imbalances between buyers and sellers that affects a shape of price trajectories. Moreover, the paper showed that such system can converge to two qualitatively different states, one with a zero (and fixed) terminal state and the other with a strictly positive (though volatile) terminal state.

  While the foundational work laid crucial groundwork by linking aggregate money supply and market imbalances to the direction and shape of price trajectories, it did not explicitly model the dynamic shifts between the qualitatively distinct regimes culminating in either a zero or a strictly positive terminal price. This limitation motivates the current paper, which seeks to unravel the hidden dynamics driving these transitions.

  Both papers share the ASME framework ot the artificial, order-driven stock market. Autonomous agents or traders who continuously observe market dynamics, interact through a limit order book and respond by submitting ask and bid orders according to their preferences, resource endowments, personal beliefs, trading objectives and market conditions with a prime objective to accummulate wealth. The conceptualization of the stock market as a multi-agent system has deep roots in both economic theory and computational modeling. Early contributions to this field include seminal works such as \cite{kim1989investment}, who explored portfolio optimization under uncertainty, laying the groundwork for understanding how agents allocate resources based on risk-return trade-offs. Similarly, \cite{levy1994microscopic} introduced a microscopic simulation framework to study wealth distribution dynamics among traders, highlighting the role of heterogeneity in shaping market outcomes. Building on these foundations, \cite{palmer1999artificial} developed artificial stock market models to simulate the behavior of interacting agents, demonstrating how aggregate market properties emerge from localized interactions. Other notable contributions include \cite{lebaron2002building}, who advanced agent-based modeling techniques to incorporate learning and adaptation, and \cite{lux1999scaling}, whose work elucidated the scaling properties of financial markets through the lens of herding behavior and speculative dynamics.
  
  However, unlike its conceptual ancestor, this paper explicitly studies the bifurcative nature of the ASME trading framework. Nonlinear systems can exhibit sensitive dependence on initial conditions or stable/unstable regions depending on external conditions or system parameters. Simulations have shown that the system exhibits bistability, with trajectories converging to two distinct types of end states: (1) a single, deterministic equilibrium where prices converge to zero and remain there and (2) a stochastic equilibrium where prices converge to a normal distribution around a positive mean. The paper shows that there exist conditions that induce bistability even under the same initial conditions. Trajectories eventually settle near one of two distinct price levels with noise causing small deviations that push different trajectories into different basins of attraction. This is the hallmark of stochastic bistability arising from a bifurcation (e.g. saddle-node or supercritical pitchfork) in the underlying deterministic dynamics.

  The system resembles a supercritical pitchfork bifurcation, which occurs at a critical money threshold that is proportional to the number of shares. It is the aim of the paper to study bifurcation properties of the model and their stability. The key goal is to understand the dynamics of transitions in the market that is studied with the Hidden Markov Model (HMM). The choice of an HMM framework is predicated on its inherent ability to model systems characterized by unobserved, underlying states that govern the observed sequence of market activities, specifically price changes and bid-ask imbalance. By assuming that the artificial stock market operates within a finite set of such hidden states, each with its own probabilistic characteristics for generating the observed data, we aim to identify and characterize the distinct dynamical regimes that the market traverses. This approach allows us to move beyond simply observing the bistable outcomes and delve into the temporal dependencies and the probabilistic nature of the market's evolution, particularly in the vicinity of the bifurcation point identified through the lens of bifurcation theory. HMM has been used for an identification of patterns in financial markets, for instance, \cite{aydinhan2024identifying}.

  The contributions of this paper are manifold. The core finding of intrinsic bistability in an artificial stock market with homogeneous agents is significant. It challenges the notion that complex market behaviors necessarily arise from heterogeneous agent characteristics or external shocks. The paper effectively argues that the micro-level trading rules themselves can generate divergent market outcomes. It, namely, shows that even identical initial conditions can lead the system to qualitatively different long-run price equilibria: a deterministic zero-price state and a persistent positive-price equilibrium which underscores the nonlinear and path-dependent nature of price evolution in the model. The attractor's topological structure, as visualized through the clustering and separation, confirms that the agent-based price dynamics possess intrinsic bistability, which is not a result of randomness but an emergent property of micro-level trading rules.
  
  Secondly, attractors are separated with a metastable region, functioning as a transient phase space, where trajectories exhibit heightened sensitivity to stochastic perturbations. The trajectories converging toward $P^* = 0$ exhibit protracted transitions and greater variance, suggesting that the path to the terminal, absorbing state is dynamically extended. The possibility of the price system to converge to the $P^* = 0$ pose implications that are beyond the paper, see, for instance, \cite{battiston2016complexity}.
  
  Thirdly, by explicitly modeling the transitions between these regimes, the paper aims to enhance our understanding of the complex dynamics underlying the market's behavior near the bifurcation point. The fractal dimension values observed indicate that the dynamics of the artificial market are nonlinear but not fully chaotic. The system appears to exhibit intermittent chaotic behavior or low-dimensional chaos.
  
  Fourthly, it offers a data-driven characterization of the market's hidden dynamical regimes, linking these regimes to the previously identified zero and positive price equilibria. Finally, the inclusion of bid-ask imbalance as an observed variable provides a richer perspective on the market's microstructure and its role in influencing the transitions between different dynamical states.

  The paper is organized as follows. Section 2 summarizes literature on the topic with a special focus on modeling the LOB. Mathematical model is developed in Section 3. Trader's decision-making, despite being myopic, random, stems from solving Bellman equation, whose closed form solution defines traders' optimal strategy. If his expectation of price change is positive, he submits a buy order, and a sell order in case of a drop. Section 4 presents a summary of the simulation results. This section also includes various empirical tests designed to analyze the bifurcative and chaotic characteristics of the system. The simulation results indicate a bimodal state, often preceded by a transient phase. Section 5 then analyzes the probability of price trajectories reaching the zero-price state. Section 6 frames the system within the context of a multivariate Hidden Markov Model with two hidden states. Finally, Section 7 provides concluding remarks.


  \section{Related Works}
  Our conceptualization of the artificial stock market exchange (ASME) is rooted in the fundamental functionality of a limit order book (LOB) market, a market microstructure that has evolved significantly with technological advancements \cite{Garman1976, madhavan2000market, o2002market, ohara2015high, hasbrouck2007empirical}. Early works on LOBs explored the role of market makers and the bid-ask spread as a cost of immediacy \cite{demsetz1968cost, amihud1986asset}, as well as informational disadvantages for early traders \cite{copeland1983information}.

  Early theoretical models often employed continuous-time double auction frameworks to analyze price formation, with Kyle's model \cite{kyle1985continuous} being foundational for understanding information incorporation through market orders. The role of asymmetric information and adverse selection in spread determination was emphasized by \cite{glosten1985bid, glosten1994electronic}, further developed by studies on optimal order timing \cite{admati1988theory} and stealth trading \cite{barclay1993stealth, chakravarty2001stealth}. Auction theory also provided insights into the emergence of bid-ask spreads through the analysis of bidding behavior \cite{wilson1979auctions, back1993auctions, milgrom1982theory}.
  
  More formal modeling of limit orders often involves solving an investor's utility maximization problem in continuous time \cite{parlour2008limit}. Static equilibrium LOB models \cite{Rock1996, glosten1994electronic, chakravarty1995integrated, handa1996limit, seppi1997liquidity, foucault2008competition} provided initial analytical frameworks. Recent research increasingly models limit order markets as sequential games, focusing on the strategic interaction of agents submitting orders \cite{foucault1999order, parlour1998price, foucault2005limit, goettler2005equilibrium, goettler2009informed, rocsu2009dynamic}.
  
  Another significant approach involves modeling the arrival and cancellation of orders using stochastic processes, with Poisson processes \cite{cont2003financial, cont2010stochastic, cont2013price} and Hawkes processes \cite{hawkes1971spectra, lu2018high, bacry2015} being prominent examples. More recently, machine learning techniques like deep neural networks and GANs have been applied to LOB modeling \cite{sirignano2019deep, zhang2019deeplob, prenzel2022dynamic}. Queuing theory offers a natural framework for analyzing order flow dynamics, focusing on waiting times and queue lengths \cite{gross2008fundamentals, weber2005order, huang2015simulating, daniels2003quantitative, cont2013price}. These stochastic models and queuing theory approaches often aim to capture aggregate trader behavior for analytical tractability \cite{gueant2016financial}. Comprehensive surveys of LOB modeling can be found in \cite{parlour2008limit}, \cite{abergel2016limit}, \cite{gould2013limit}.
  
  In parallel, agent-based models (ABMs) offer a bottom-up perspective by simulating the interactions of heterogeneous agents with predefined strategies \cite{tesfatsion2006agent}. Early influential ABMs \cite{bak1997price}, \cite{palmer1999artificial}, \cite{donangelo1998money}, \cite{donangelo1997statistical}, \cite{kim1993stock}, \cite{levy1994microscopic}, \cite{levy1995phase}, \cite{levy1995dynamical}, \cite{levy1996market}, \cite{levy2000new}, \cite{levy2000microscopic}, \cite{levy1997towards}, \cite{huang2001power}, \cite{hommes2006heterogeneous}, \cite{hommes2017booms}, \cite{solomon1999money}, \cite{lux1999scaling}, \cite{lux2000volatility}, \cite{boswijk2007behavioral}, \cite{lebaron2021microconsistency}, \cite{fievet2018calibrating} demonstrated the emergence of complex market phenomena from local interactions.
  
  Specifically relevant to LOB modeling is the work by \cite{cont2000scaling} and subsequent models \cite{bouchaud2002statistical}, \cite{chiarella2002simulation}, \cite{preis2006multi}, \cite{farmer2005predictive}, \cite{thurner2012leverage}, \cite{wang2021spoofing}, \cite{coletta2022learning} that simulate order book dynamics and the interplay of market and limit orders. These models highlight the importance of agent interaction and feedback loops in shaping LOB behavior \cite{challet2004minority, nevmyvaka2006reinforcement, lebaron2006agent}. Surveys of agent-based computational finance include \cite{samanidou2007agent}, \cite{steinbacher2021advances}, \cite{axtell2025agent}.

  Complexity of financial systems as a bifurcative and chaotic system has been studied, for instance by \cite{gao2009chaos}, \cite{yu2012dynamic}, \cite{hsieh1991chaos}, \cite{anufriev2020chaos}, \cite{gardini20232d}, \cite{panchuk2018financial}. However, this kind of research that would focus on the microstructure of the LOB, as is the case in this paper, is scant if existent at all.

  \section{The Model}
  The model is the same as in \cite{steinbacher2025mathematical}. It is a discrete-time, discrete-state model of interacting agents who trade a single asset $S$ and whose wealth in time $t \in \{0, 1, ..., T\}$, $W_t^i = M_t^i + S_t^i P_t, \quad i \in \{1, 2, ..., N\}$, is composed of monetary holdings $M_t^i$ and holdings of the asset $S_t^i$ with $P_t \in \mathbb{R}^+$ being the market price of the asset at time $t$.\footnote{Our earlier paper on the topic \cite{steinbacher2025mathematical} provides a comprehensive derivation and analysis of the model.} Since short selling through negative money holdings is not allowed and $S_t^i \in \mathbb{Z}^+$ which means that only non-negative integer position in the asset is allowed.

  \subsection{Agents}

  Agents in the model trade the stock by maximizing their expected wealth of the form 
  \begin{equation}
    \max_{\{\pi_t^i\}_{t=0}^{T-1}} \mathbb{E}_0\left[\sum_{t=1}^{T} \beta^{t-1} u(W_t^i)\right],
    \label{eq:objective}
  \end{equation}
  where $\pi_t^i$ represents the trading strategy of agent $i$ at time $t$; $\beta \in (0, 1)$ is the discount factor, reflecting the time preference of the agent; $u: \mathbb{R} \rightarrow \mathbb{R}$ is a utility function, CRRA in our case, which is strictly increasing, concave, and twice continuously differentiable of the form: $u(W) = \frac{W^{1-\gamma}}{1-\gamma}$ for $\gamma \neq 1$ and $u(W) = \ln(W)$ for $\gamma = 1$, where $\gamma > 0$ is the coefficient of relative risk aversion; $\mathbb{E}_0[\cdot]$ denotes the expectation operator conditional on the information available at time $t=0$ and $t=1,2,...,T$ is time domain.

  Solving the utility function for the price $P$ makes an optimal strategy for a trader that gets the form. 
  \begin{align}
    \label{eq:optimal_strategy}
        A_{t+1} &= 
        \begin{cases} 
            Bid & P_t^{\ast} < E[P_{t+1}] - \frac{\gamma}{2} \frac{\operatorname{Var}(P_{t+1})}{W^*}, \\
            Ask & P_t^{\ast} > E[P_{t+1}] + \frac{\gamma}{2} \frac{\operatorname{Var}(P_{t+1})}{W^*}, \\
            \text{Nothing} & \text{else},
        \end{cases}
  \end{align}

  Equation \ref{eq:optimal_strategy} states that a trader should follow his risk-adjusted pricing. Given that a myopic, random trader does not incorporate risk consideration into the trading decision,\footnote{Effectively sets $\gamma = 0$.} this can be simplified to
    
    \begin{align}
      \label{eq:agent_choice}
      A_{t+1} &= 
      \begin{cases} 
            Bid & P_t^{\ast} < E[P_{t+1}], \\
            Ask & P_t^{\ast} > E[P_{t+1}], \\
            \text{Nothing} & \text{else},
      \end{cases}
  \end{align}
    
  This means that the trader submits an ask order if his private expectation of the price-change is a drop and a bid order if the price is expected to rise. A trader $i$ predicts the future price as
  \begin{equation}
    \label{eq:expected_price}
    E_t^i[P_{t+1} | P_t] = P_t \cdot (1 + \Delta_i),
  \end{equation}
  where $\Delta_i \sim U(-\sigma,\sigma) $ is independently and identically distributed (i.i.d.) from a uniform distribution $U(-\sigma, \sigma)$.

  \subsection{Limit Order Book}
  Agents submit their limit ask and bid orders to the order book. Conceptually, our definition of the limit order is consistent with the \cite{parlour2008limit} where it is defined as an ex ante pre-commitment made on date $t$ to trade up to a given amount $x$ of a security $j$ at a pre-specified limit price $p$ that is in force until filled or cancelled. An orderbook is a collection of bids and asks. At any given time $t$ a set of bids and asks can respectively be defined as

  \begin{definition}[Order Book]
      \label{def:order_book}
      The order book is a collection of all bids $B_t$ and asks $A_t$ in time $t$, where:
      \begin{itemize}
          \item $B_t = \{ (P^i_b, q^i_b) : P_b \in \mathbb{R}^+, q_b \in \mathbb{N}, i \in \{1, 2, ..., N\} \}$: The set of all bid orders at time $t$. Each bid $b \in B_t$ is a tuple $b^i = (p_b, q_b, t_b)$, where $p_b \in \mathbb{R}^+$ is the bid price, $q_b \in \mathbb{N}$ is the bid quantity, and $t_b$ is the timestamp of the bid order.
          \item $A_t = \{ (P^i_a, q^i_a) : P_a \in \mathbb{R}^+, q_a \in \mathbb{N}, i \in \{1, 2, ..., N\} \}$: The set of all ask orders at time $t$. Each ask $a \in A_t$ is a tuple $a^i = (p_a, q_a, t_a)$, where $p_a \in \mathbb{R}^+$ is the ask price, $q_a \in \mathbb{N}$ is the ask quantity, and $t_a$ is the timestamp of the ask order.
      \end{itemize}
  \end{definition}

  The market clearing process determines which orders are matched and executed, establishing the transaction price and quantities. We assume a price priority matching rule. This means that all bids are ranked in descending order of price and asks are ranked in ascending order of price.

A trade is a pair of the best bid and the best ask prices. The best bid price ($P_t^b$) is the highest bid price available in the market at time $t$. The best ask price ($P_t^a$) is the lowest ask price available at time $t$. We define them as:

        \[
        P_t^b = \sup \{ P_b : (P^i_b, q^i_b) \in B_t \}
        \]
        \[
        P_t^a = \inf \{ P_a : (P^i_a, q^i_a) \in A_t \}
        \]

Where $\sup$ denotes the supremum or least upper bound and $\inf$ denotes the infimum that is greatest lower bound. A trade is defined as a matched pair of a bid and an ask order and can be formally writen as:

\begin{definition}[]
    \label{def:trade}
    A trade $T$ is a tuple $T = (b, a, p, q, t)$, where:
    \begin{itemize}
        \item $b$ is a bid order.
        \item $a$ is an ask order.
        \item $p \in \mathbb{R}^+$ is the transaction price, ie. the price at which the trade is executed.
        \item $q \in \mathbb{N}$ is the traded quantity, ie. the number of shares or contracts exchanged.
        \item $t$ is the transaction timestamp, ie. the time at which the trade is executed.
    \end{itemize}
\end{definition}

A trade $T$ occurs when a bid $b \in B_t$ and an ask $a \in A_t$ iff $p_b \geq p_a$ and $q = \min(q_b, q_a)$. The former can be defined as a matching condition that states that the bid price must be greater than or equal to the ask price. The later is a quantity condition saying that the traded quantity is the minimum of the bid and ask quantities.

The resulting price of the trade lies between the bid and ask prices or $p_a \le p \le p_b$. In general, it is common to opt for the mid-price or $p = \frac{p_b + p_a}{2}$ which is the stance taken in the paper as well.

\subsection{Solving for Price Change}
It was shown in the \cite{steinbacher2025mathematical} that the approximate expected price change equals

\begin{equation}
  \label{eq:expected_price_change}
    E[\Delta P_{t+1}^{(approx)} | P_t = p] \approx \frac{\sigma}{2} P_t \left( \frac{1}{N_A+1} - \frac{1}{N_B+1} \right),
\end{equation}

with $N_A$ number of sellers, $N_B$ number of buyers, $P_t$ current price and $\sigma$ parameter of the trader's expected price change from Equation \ref{eq:expected_price}. Equation \ref{eq:expected_price_change}, basically, states that the direction of the price change is defined by the market imbalance between buyers and sellers, pushing the price downwards if the ratio is in favor of the former or upwards in the case of the latter.

  \section{Simulations}
  \cite{steinbacher2025mathematical} demonstrated that variation in the control parameter $M$, ceteris paribus, can lead to the emergence of two qualitatively distinct attractors: a trivial fixed point $P^*=0$ and a nontrivial state $P^*>0$. This observation suggests that the system undergoes a bifurcation, wherein a marginal change in initial conditions, like $M$, results in a qualitative shift in long-term dynamics. Such behavior is indicative of bistability, where the system admits two coexisting stable equilibria, and small perturbations or differences in initial conditions may determine which attractor the system converges to.

  Numerical simulations further reveal that this kind of bistability, manifesting as convergence to either $P^*=0$ or $P^*>0$, can occur even for identical initial conditions. In particular, there exists a continuous region in parameter space, $M \in (M_1,M_2)$, ceteris paribus, across which bistability persists. This region forms a transition band separating domains of monostability at either $P^*(M<M_1)=0$ or $P^*(M>M_2)>0$. While the current paper does not explore the entire domain of this bistable regime, it focuses on one such representative example that illustrates the existence and nature of this phenomenon.

  \subsection{Simulation Parameters}

  To validate and explore the theoretical model and the emergent price trajectory, we conduct a series of simulation batches. The general simulation setup is configured as follows. The agent population is set to $N = 10,000$ traders, each initialized with a wealth endowment $W_0^i(M^i,S^i)$ comprising of monetary holdings $M^i$ and stock holdings $S^i$. Initial holdings of wealth are equal for all traders, that is $M_0^i = M_0 \forall i$ and $S_0^i = S_0 \forall i$ in every simulation run. Stocks are traded in units and cannot be shorted and traders are under a hard-budget constraint, forbidding them to trade on a margin. Money does not earn any interest.

  The trading horizon spans for $t=1,...,5,000$ discrete time periods. In each iteration $t$, a myopic, random (zero intelligent)\footnote{Traders following zero intelligence (ZI) strategy by placing random order submissions have been widely used for the study of LOB, for instance, \cite{gode1993allocative} or \cite{farmer2005predictive}.} trader makes a trading decision based on Equation \ref{eq:agent_choice} where $E(P_{t+1} | I_t, P_t) = P_t \cdot (1+ \Delta \sim \mathbb{U}(-0.15,0.15))$ with $\Delta$ a uniform random variable on the interval. If the price is expected to increase by the trader $i$, a bid (buy) order is submitted. Alternativelly, if the trader expects a fall of the price, he submits an ask (sell) order. Both bid and ask orders are submitted to the orderbook for 1 unit stock $S$ at the beginning of the time interval $t$. Once submitted orders can not be modified. The orderbook, as a match-maker, iteratively matches best bids with best asks, that is the highest-priced bid and the lowest-priced ask, until such pairs can be formed and settles them at the middle of the bid's and the ask's price. Any unmatched order at the end of the time interval $t$ is cancelled and discarded. Trading does not involve any costs. Settlements are promptly done. A settled price of the last trade of an iteration is recorded as the closed price of the iteration $P_t$ that becomes public knowledge available to all traders.

  The initial stock price is set at $P_0 = 10.0$ and each trader is initially endowed with $S_0=20$ stocks. Initial stock endowment and initial stock price are held constant during all repetitions of all batches. The initial monetary holdings $M_0=174$.\footnote{The source code of the artificial stock-market environment was written in C++ and was compiled for the Xcode. In the context of an object-oriented framework of the C++, a trader is modeled as an object with a lifetime and a scope that encapsulates both its state (private and public data) and its behavior (actions). In essence, the C++ trader object acts as a software representation of an autonomous agent operating within the simulated stock market, with its own internal characteristics and a defined set of actions it can take to interact with the market and other agents.}

  \subsection{Simulation Results}

\begin{figure}[h]
    \centering
    \includegraphics[width=\textwidth]{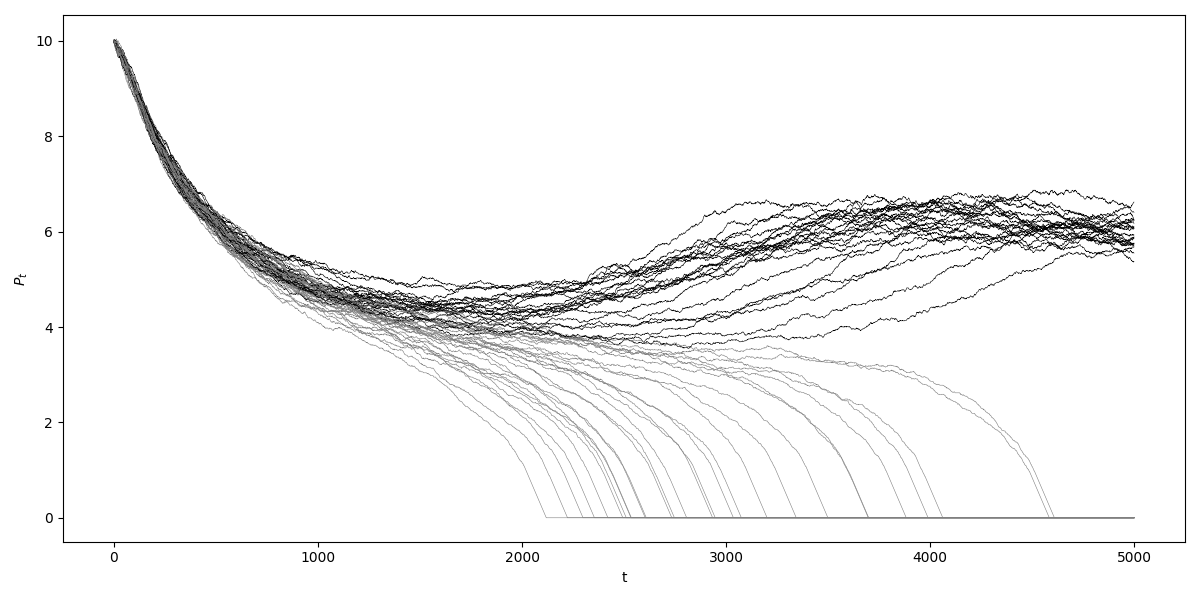}
    \caption{Simulation Results}
    \label{fig:base_simulations}
  \end{figure}

  Simulation results of 50 independent repetitions all starting from the same initial conditions are shown in Figure \ref{fig:base_simulations}. 

  Despite the same initial conditions, the trajectories fan out into different terminal states, with 22 of them settling at higher values $P_T>0$ and remaining 28 trajectories dropping towards $P_T=0$. Such state strongly indicates that (1) small stochastic variations (noise) in the system's evolution are pushing trajectories onto different basins of attraction and (2) if the boundary separating the two attractors is close to the initial condition (or is "thin"), even tiny random perturbations can send a trajectory to one attractor or the other, that is noise-sensitive basin boundaries.

  First impression is that trajectories settle (or nearly settle) into a stable level rather than exploding or oscillating indefinitely making a case for the a bistability under some stochastic influence rather than a typical chaos which would show sensitive dependence with ongoing, non-convergent fluctuations.

  \subsubsection{Basic Analysis}
  \paragraph{Test of Stationarity}
  Stationarity of trajectories was assessed using the Augmented Dickey-Fuller (ADF) unit root test, see, for instance, \cite{hamilton2020time}. The ADF test is based on fitting a model of the form:
  
  \begin{equation}
    \Delta y_t = \alpha + \beta t + \gamma y_{t-1} + \sum_{i=1}^{p-1} \delta_i \Delta y_{t-i} + \epsilon_t,    
  \end{equation}
  
  where $\Delta y_t = y_t - y_{t-1}$, and the test examines the null hypothesis $H_0: \gamma = 0$ (presence of a unit root, i.e., non-stationarity) against the alternative hypothesis $H_1: \gamma < 0$ (stationarity).

  \begin{table}[]
    \centering
    \caption{ADF Test: $P^*>0$}
    \label{tbl:adf_positive}
      \begin{tabular}{l|c|c}
        \toprule
                   &  ADF    & p-value\\
        \midrule
               AVG & -5.1665 & 0.0006 \\
        \hline MIN & -7.6685 & 0.0000 \\
        \hline MAX & -3.5092 & 0.0078 \\
        \hline STD & 1.1022  & 0.0017 \\
        \hline N   & 22      & 22     \\
        \bottomrule
      \end{tabular}
  \end{table}

  The results of the ADF test for trajectories with $P^*>0$ are presented in Table \ref{tbl:adf_positive}. The table shows the ADF test statistic and the corresponding p-value. For all statistical measures of the trajectories, the low p-values ($p < 0.05$) lead to the rejection of the null hypothesis, providing strong evidence that the trajectories of the $P^*>0$ group exhibit stationarity in the weak sense, meaning their first two moments are time-invariant ($E[y_t] = \mu$, $Cov(y_t, y_{t-k}) = \gamma(k)$). This reinforces the interpretation that these trajectories converge to a stable stochastic equilibrium.

  \paragraph{First and Second Moments}

  Let $x_{i,t}$ represent the price of the trajectory at time $t \in \{1, 2, ..., T\}$ in the $i$-th repetition, where $i \in \{1, 2, ..., R\}$. The average price trajectory as a function of time, conditioned on the terminal price $P^*$, is denoted by $X(\bar{x}_t | P^*)$, where the temporal average at each time point $t$ is given by $\bar{x}_t = \frac{1}{R} \sum_{i=1}^{R} x_{i,t}$. Similarly, the temporal standard deviation of the price trajectory, conditioned on $P^*$, is denoted by $\sigma(\sigma_t | P^*)$, where the standard deviation at each time point $t$ is calculated as $\sigma_t = \sqrt{\frac{1}{R-1} \sum_{i=1}^{R} (x_{i,t} - \bar{x}_t)^2}.$

  \begin{figure}[h]
    \centering
    \includegraphics[width=\textwidth]{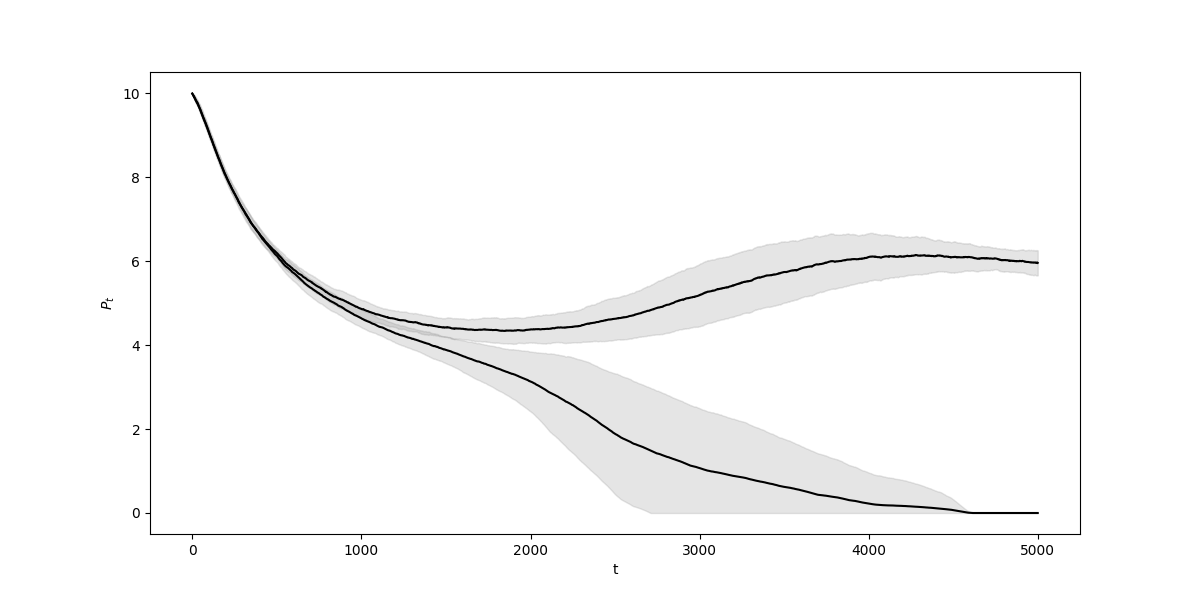}
    \caption{Mean Price Trajectories with Standard Deviation}
    \label{fig:base_simulations_mean_std}
  \end{figure}

  Figure \ref{fig:base_simulations_mean_std} depicts the temporal evolution of the average price trajectories, $X(\bar{x}_t | P^*)$, categorized into two distinct groups based on their terminal price $P^* = 0$ and $P^* > 0$. The shaded areas surrounding each average trajectory represent one standard deviation, $\sigma(\sigma_t | P^*)$, at each time point, illustrating the volatility or dispersion of individual trajectories within each respective group. The figure provides a clear visual representation of the bistable nature of the artificial stock market model, demonstrating how initially similar trajectories generated under the same initial conditions diverge over time to converge towards either a zero or a positive price equilibrium, each characterized by varying degrees of volatility.
 
  Periods in the neighborhood of the divergence point, say for $t \in [t_d - \delta_1, t_d + \delta_2]$ where $\delta_1, \delta_2 > 0$, exhibit higher volatility, i.e., larger values of $\sigma_t$ for both groups. This suggests that during this sensitive phase, the market's response to stochastic noise is amplified, playing a crucial role in determining the final equilibrium state. The group of trajectories converging to $P^*=0$ shows relatively larger volatility during the transient phase ($t < t_c$, where $t_c$ denotes the approximate convergence time), such that $\sigma_{t|P^*=0} > \sigma_{t|P^*>0}$ for $t < t_c$. This indicates a greater dispersion among individual trajectories as they approach the zero equilibrium.

  Convergence for the zero price trajectories is considered complete when, for all repetitions $i$, $x_{i,t} \rightarrow 0$ as $t \rightarrow T$, and consequently, the standard deviation $\sigma_{t|P^*=0} \rightarrow 0$ as $t \rightarrow T$. However, for the positive-priced group, the volatility around the positive mean persists until the end of the observation period, with $\sigma_{T|P^*>0} > \epsilon$, with $\epsilon > 0$. This non-zero terminal volatility suggests that the equilibrium at $P^*>0$ is stochastic, characterized by ongoing price fluctuations and inherent randomness.

  The diverging ensemble averages of price trajectories closely mirrors the structure of a supercritical pitchfork bifurcation in a stochastic dynamical system \cite{zwillinger2021handbook,kuznetsov1998elements}. Initially symmetric decay gives way to a regime split—where some trajectories collapse to zero while others converge to a metastable, fluctuating high-price state. This pattern suggests an underlying symmetry-breaking transition, likely triggered by nonlinear interactions among agents modulated by stochasticity.


  \subsubsection{Analysis of Bifurcative Characteristics}
  To understand the bifurcating behavior of the simulated price trajectories, this section, first, estimates the separatrix point, second, calculates time to zero for the $P^*=0$ trajectories and, third, studies Poincaré maps. These methods provide complementary insights into the system's dynamical behavior. \cite{kuznetsov1998elements} or \cite{zwillinger2021handbook} are comprehensive sources that cover material on dynamical systems used within the section.

  \paragraph{Separatrix}

  Estimations suggest that the separation between both groups, the separatrix price $P_c$,\footnote{A separatrix is a boundary in the phase space of a dynamical system that separates regions of qualitatively different asymptotic behavior.} appeared on the range between $3.54$ (entropy-based method) to $3.733$ (SVM), with the classification-based method yielding a value of $3.665$.\footnote{The detailed procedures for each algorithm are provided in the Appendix.} Additionally, time-dependent tests, performed between the high-price and low-price groups show that the statistically significant divergence of trajectories of both groups started within $t=(650,...,700)$, when resulting p-values of a T-test for the means of two independent samples of scores and a Mann-Whitney U test performed over an $P_{t,avg} = \left(\sum_{t-10}^{t}P_t \right) / 10$ for $t=10,...,T$ between the high-price and low-price groups start rejecting the null hypotheses of distribution equallity (the Mann-Whitney U test) and equality of means (the T-test). For instance, $U_{(t=650)}=202.5000(0.0401)$ and $\text{T-Test}_{(t=650)}=-2.2845(0.0268)$ with p-values in the brackets.

  The price dispersion of the low-priced trajectories at the equilibrium price $P_T=0$ is zero, which implies that the escape rate of the equilibrium is zero. Terminal prices of the high-priced state do not converge to a single price but show a certain level of oscillation, given summary statistics, like $\overline{P}_{5000}=5.9577$, $Med(P_{5000})=5.91$, $\sigma_{P_{5000}}=0.2887$, $min(P_{5000})=5.37$, $max(P_{5000})=6.62$. A failure to reject the Anderson-Darling test for normality $AD=0.24275$ indicates that the distribution of terminal prices of the group is consistent with the normal distribution. As a corollary, the equilibrium is not a single, fixed price point. Instead, it is a stable state where prices oscillate randomly around a mean value.

  \paragraph{Time to Zero}
  As shown in Table \ref{tbl:po_time2zero}, the distribution of the time to zero for these trajectories when $P^*=0$ is centered around $t=3053.5$, with a median of $t=2868.5$. Time to zero is set to the earliest time step at which the trajectory reached zero. There is a moderate level of variability in the time to zero, as indicated by the standard deviation and coefficient of variation. The positive skewness suggests that while most trajectories reach zero relatively quickly, some take considerably longer. The negative excess kurtosis indicates a flatter distribution with less pronounced tails compared to a normal distribution. The range of times to zero is quite broad, from $t=2117$ to $t=4607$. The IQR further clarifies that the middle 50\% of the data falls within a range of $t=1130.8$.

  \begin{table}[!htp]
    \centering
    \caption{$P^*=0$: Time to Zero}
    \label{tbl:po_time2zero}
      \begin{tabular}{lc|lc}
        \toprule
               Mean     &   3053.5 & Skew     &  0.76736 \\
        \hline Median   &   2868.5 & Ex. kurt & -0.42302 \\
        \hline Minimum  &   2117.0 & 5\%      &   2163.8 \\
        \hline Maximum  &   4607.0 & 95\%     &   4595.8 \\
        \hline STD      &   702.58 & IQR      &   1130.8 \\
        \hline C.V.     &   0.2301 & N        &       28 \\
      \bottomrule
      \end{tabular}
  \end{table}

  \paragraph{Poincaré Maps}
  This section studies the stochasticity of the system using Poincaré maps. The Poincaré map is a technique employed in the analysis of dynamical systems, focusing on the intersections of the system's trajectory in the phase space with a lower-dimensional subspace, referred to as the Poincaré section. This method effectively reduces the continuous-time dynamics to a discrete-time map, enabling the examination of the system's long-term behavior and stability.

  The phase space is a multidimensional space where each point $\mathbf{y}_t$ represents the state of the system at a particular time $t$. For a scalar time series $\{x_t\}_{t=1}^N$, the method of time delays allows for the reconstruction of the phase space. An embedded vector in this reconstructed phase space at time $t'$ can be defined as:
  \[
  \mathbf{y}_{t'} = (x_{t'}, x_{t' - \tau}, ..., x_{t' - (m-1)\tau}) \in \mathbb{R}^m
  \]
  where $m$ is the embedding dimension, which determines the number of past values of the time series used to define each point in this reconstructed phase space, and $\tau$ is the time delay, representing the lag between the components of the embedded vectors. The embedding dimension $m$ dictates the dimensionality of the reconstructed phase space, while the time delay $\tau$ governs the temporal separation between the components of the state vector.

\begin{figure}[h]
  \centering
  \includegraphics[width=\textwidth]{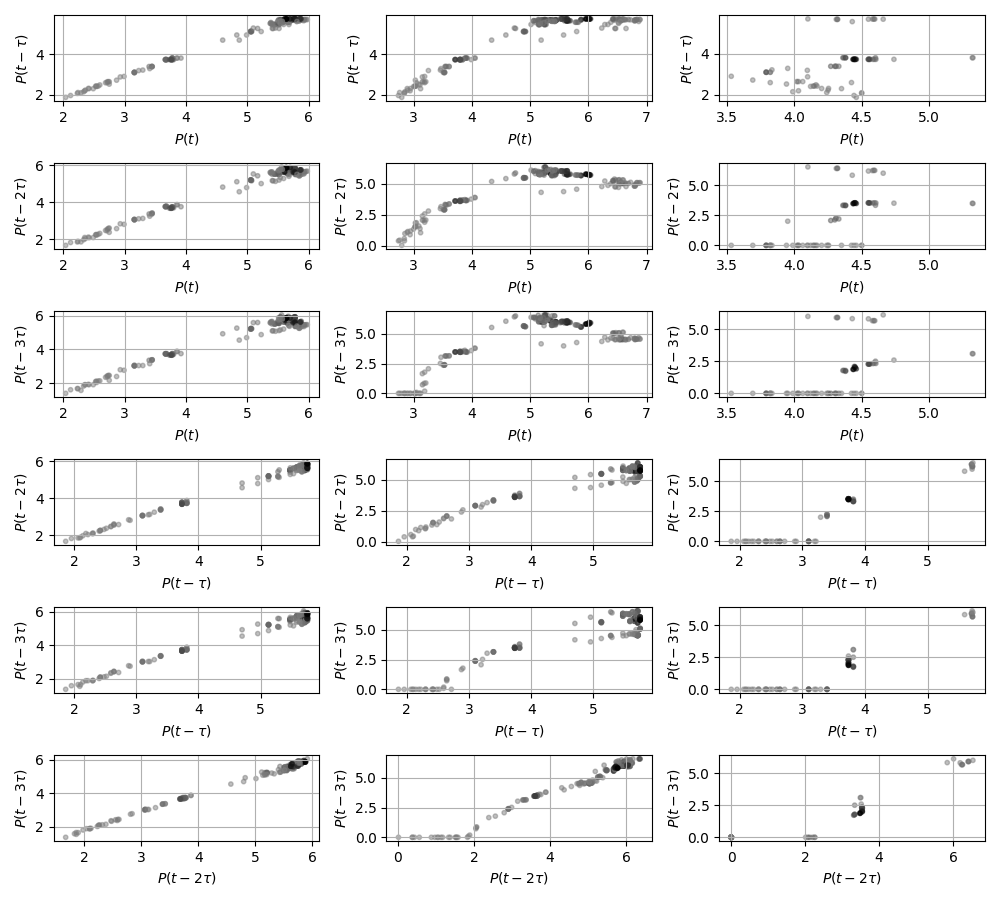}
  \caption{Column-Wise Poincaré Maps from Price Series}
  \label{fig:poincare_maps}
\end{figure}
  
  Figure \ref{fig:poincare_maps} shows a set of Poincaré maps for a fixed embedding dimension $m=4$ arranged column-wise  for three different time delays $\tau = 50, 250, 1000$ that let us perform a more granular, high-dimensional phase space analysis. To construct these maps, a suitable Poincaré section (a subspace of dimension $m-1$) is chosen, and the points where the reconstructed trajectory intersects this section (typically with a specific direction of crossing) are plotted. The patterns observed in these Poincaré maps can provide insights into the nature of the system's dynamics; for instance, densely scattered points might suggest stochastic behavior, while more structured patterns could indicate deterministic or chaotic dynamics.
  
  For $\tau=50$ (first column), the points in the Poincaré section exhibit a strong linear correlation along the diagonal in all projections, indicating a high degree of autocorrelation between $x_t$ and $x_{t-50}$. The minimal spread around this linear trend suggests low variance. The dominant attractor corresponds to the high-price regime ($P \approx 5.5$), evidenced by the high density of points in that region, while the low-price attractor ($P \approx 0$) is sparsely represented. The reconstructed phase space appears to be a near one-dimensional projection, implying that the chosen embedding parameters do not adequately unfold the system's dynamics. The short time delay $\tau=50$ is insufficient to effectively distinguish independent states of the system.

  Setting $\tau=250$ (second column) provides a more intricate and potentially more informative representation of the system's dynamics, revealing patterns consistent with chaotic behavior. The non-uniform distribution of points across the projections indicates deterministic dynamics. The clustering of points in specific regions signifies the presence of attractors, which are regions in the phase space towards which trajectories asymptotically evolve. Two primary regions of high point density are discernible, suggesting the existence of at least two attractors. These clusters are separated by regions with a lower point density, implying less frequent or rapid transitions between the associated dynamical regimes. The smooth variation of point density within each cluster and along the transitional areas suggests a continuous evolution of the system's state. The observation of multiple distinct regions of attraction, separated by less frequently visited areas, is a hallmark of multistability. The potential for the system to reside for extended periods near each of these clusters before transitioning suggests a metastable regime. This setting seems optimal for embedding since time delays are sufficiently decorrelated to unfold the manifold, but not too long to be overwhelmed by stochasticity. The attractor appears nonlinear but structured, suggesting a low- to moderate-dimensional chaotic system.
  
  For $\tau=1000$ (third column), the price values in the embedded vectors are separated by a relatively long interval that has exceeded the system's decorrelation time. This choice of $\tau$ effectively isolates the basins of attraction but may obscure finer dynamical details. The Poincaré map clearly reveals two dense and stable basins of attraction corresponding to the equilibria at $P^*=0$ and $P^* > 0$. These basins are separated by sparse or transient regions that are visited infrequently and non-periodically, which is indicative of a metastable regime, particularly around $P_t \approx 3.8$, a value close to the calculated separatrix. The distinct clustering of intersection points for $\tau=1000$ highlights the specific regions in the phase space that the system preferentially visits. The sparser distribution and the clearer separation of clusters compared to smaller time delays suggest that this longer time delay captures longer-term behaviors and a less frequent crossing of the Poincaré section.

  The results clearly show that only with sufficiently long delay ($\tau=1000$) does the embedding reveal the true bifurcative nature of the system. The attractor's topological structure, as visualized through the clustering and separation, confirms that the agent-based price dynamics possess intrinsic bistability, which is not a result of randomness but an emergent property of micro-level trading rules. 

  The evolution of the system with two diverging average price trajectories over time strongly resembles a supercritical pitchfork bifurcation in a stochastic setting. In classical dynamical systems, a supercritical pitchfork bifurcation occurs when a system transitions from a single stable fixed point to two new stable fixed points and one unstable middle point as a control parameter crosses a critical value.

  \subsubsection{Chaotic Characteristics of the System}
  We study chaotic characteristics of the simulated price trajectories through three complementary measures: fractal dimension, entropy-based complexity, and the largest Lyapunov exponent. Each offers a different lens on the system's dynamical behavior.

  \paragraph{Fractal Dimension}
  The fractal (correlation) dimension quantifies the complexity of a time series by describing how the system fills its phase space. Using the Grassberger-Procaccia method (see Appendix), we compute the correlation sum for various time delays \(\tau\). Results are summarized in Table~\ref{tbl:fractal_dimension}.

  \begin{table}[h!]
  \centering
  \begin{tabular}{lccccc}
  \toprule
  \(\tau\)       & 50     & 250    & 500    & 750    & 950    \\
  \midrule
    AVG   & 0.578  & 0.779  & 0.867  & 0.793  & 0.796  \\
    STD   & 0.139  & 0.168  & 0.049  & 0.029  & 0.012  \\
    MIN   & 0.318  & 0.445  & 0.713  & 0.731  & 0.772  \\
    MAX   & 0.811  & 1.047  & 0.973  & 0.839  & 0.829  \\
  \bottomrule
  \end{tabular}
  \caption{Fractal dimension statistics across varying time delay \(\tau\)}
  \label{tbl:fractal_dimension}
  \end{table}

  The calculated fractal dimension for each trajectory consistently fell within the range of $0.8$ to $0.9$, regardless of the variations in embedding dimension and time delay parameters. This result indicates that the system exhibits a low-dimensional chaotic or quasi-periodic behavior, where the underlying dynamics are complex yet constrained within a relatively low-dimensional space. The fractal dimension values observed indicate that the dynamics of the artificial market are nonlinear but not fully chaotic. Rather, the system appears to exhibit intermittent chaotic behavior or low-dimensional chaos, where the market dynamics are influenced by a limited set of driving forces, yet still present significant irregularities and unpredictability.

  The observed fractal dimension values are lower than those typically seen in fully chaotic systems (which approach dimensions closer to $1.5-2.0$), supporting the hypothesis that the system is not fully chaotic. This could indicate that the artificial market model incorporates elements of quasi-periodic behavior or bounded chaotic dynamics, possibly reflecting real-world financial market behavior, which is often driven by both deterministic rules and random fluctuations.

  Beyond \(\tau = 500\), the average dimension stabilizes near 0.79--0.80, indicating a saturation point where further delay no longer uncovers new dynamical structure. Then \(\tau = 500\) appears to be an optimal embedding delay, balancing high fractal dimension with low variability, making it a strong candidate for optimal phase space reconstruction.

  \paragraph{Entropy-Based Complexity Analysis}
  Entropy measures were computed for 50 trajectories from the artificial limit order book stock exchange model using Approximate Entropy (ApEn), Sample Entropy (SampEn), and an estimate of the Kolmogorov-Sinai (K-S) entropy based on SampEn. The results are summarized in Table~\ref{tab:entropy}.

  \begin{table}[h!]
    \centering
    \begin{tabular}{lccc}
    \toprule
    Statistic & ApEn & SampEn & KS Est. \\
    \midrule
    Mean     & 0.0075 & 0.0063 & 0.0063 \\
    Median   & 0.0045 & 0.0040 & 0.0040 \\
    Std Dev  & 0.0051 & 0.0050 & 0.0050 \\
    Min      & 0.0023 & 0.0009 & 0.0009 \\
    Max      & 0.0168 & 0.0156 & 0.0156 \\
    25\%     & 0.0030 & 0.0016 & 0.0016 \\
    75\%     & 0.0123 & 0.0116 & 0.0116 \\
    \bottomrule
    \end{tabular}
    \caption{Entropy statistics across 50 trajectories}
    \label{tab:entropy}
    \end{table}

  Results are consistent with the those for the fractal dimension. The entropy values are uniformly low, indicating that the majority of trajectories are highly regular and predictable in most cases, which is consistent with quasi-periodic or low-dimensional deterministic dynamics. However, moderate variability in entropy values across trajectories (std $\approx$ 0.005) suggests that while most behaviors are stable, some paths exhibit higher dynamical richness which is also implied by the right-skewed distribution with a maximum SampEn of 0.0156 which implies that a subset of trajectories undergoes more irregular or transiently chaotic phases. This suggests intermittent complexity—periods of stable evolution interspersed with transient irregularities. These findings complement the fractal analysis in reinforcing the presence of low-complexity, weakly chaotic dynamics.

  \paragraph{Largest Lyapunov Exponent}
  Finally, the largest Lyapunov exponent (LLE), denoted by $\lambda$, was estimated. It measures sensitivity to initial conditions and was computed for each trajectory using the Rosenstein algorithm (see Appendix). Summary statistics are presented in Table~\ref{tbl:lyapunov_multiple}.

  \begin{table}[h!]
    \centering
    \begin{tabular}{lccccccc}
        \hline
            & ACF & $\tau=50$ & $\tau=250$ & $\tau=500$ & $\tau=1000$ & $\tau=2000$ & $\tau=4500$ \\
        \midrule
            \multicolumn{8}{c}{Total} \\
        \midrule
        $\overline{\lambda}$         & 9.25E-05 & 9.81E-04 & 2.20E-04 & 1.10E-04 & 6.00E-05 & 9.53E-06  & 4.23E-06 \\
        $\overline{\sigma}(\lambda)$ & 3.44E-05 & 2.71E-04 & 4.99E-05 & 2.78E-05 & 2.02E-05 & 1.86E-05  & 8.28E-06 \\
        $\text{min}(\lambda)$        & 3.02E-05 & 4.33E-04 & 1.30E-04 & 5.40E-05 & 1.81E-05 & -3.48E-05 & -1.54E-05 \\
        $\text{max}(\lambda)$        & 1.71E-04 & 1.49E-03 & 4.05E-04 & 2.02E-04 & 1.15E-04 & 4.63E-05  & 2.06E-05 \\
        N & 50 & 50 & 50 & 50 & 50 & 50 & 50 \\
        \midrule
            \multicolumn{8}{c}{$P^*>0$} \\
        \midrule
        $\overline{\lambda}$         & 1.20E-04 & 1.15E-03 & 2.31E-04 & 1.15E-04 & 5.22E-05 & 1.52E-05  & 6.77E-06 \\
        $\overline{\sigma}(\lambda)$ & 2.39E-05 & 1.60E-04 & 3.21E-05 & 1.60E-05 & 1.56E-05 & 1.68E-05  & 7.45E-06 \\
        $\text{min}(\lambda)$        & 8.14E-05 & 8.85E-04 & 1.77E-04 & 8.85E-05 & 1.81E-05 & -2.11E-05 & -9.36E-06 \\
        $\text{max}(\lambda)$        & 1.71E-04 & 1.49E-03 & 2.98E-04 & 1.49E-04 & 9.26E-05 & 4.63E-05  & 2.06E-05 \\
        N & 22 & 22 & 22 & 22 & 22 & 22 & 22 \\
        \midrule
            \multicolumn{8}{c}{$P^*=0$} \\
        \midrule
        $\overline{\lambda}$         & 7.08E-05 & 8.46E-04 & 2.11E-04 & 1.05E-04 & 6.61E-05 & 5.05E-06  & 2.25E-06 \\
        $\overline{\sigma}(\lambda)$ & 2.43E-05 & 2.64E-04 & 5.94E-05 & 3.40E-05 & 2.15E-05 & 1.91E-05  & 8.48E-06 \\
        $\text{min}(\lambda)$        & 3.02E-05 & 4.33E-04 & 1.30E-04 & 5.40E-05 & 2.83E-05 & -3.48E-05 & -1.54E-05 \\
        $\text{max}(\lambda)$        & 1.17E-04 & 1.34E-03 & 4.05E-04 & 2.02E-04 & 1.15E-04 & 4.35E-05  & 1.93E-05 \\
        N & 28 & 28 & 28 & 28 & 28 & 28 & 28 \\
        \hline
    \end{tabular}
    \caption{Lyapunov Exponent: Summary Statistics}
    \label{tbl:lyapunov_multiple}
\end{table}  

  The Table \ref{tbl:lyapunov_multiple} summarizes the Lyapunov exponents calculated for different groups of trajectories using various time delay ($\tau$) values. Values of $\lambda$ are consistently positive, which is a characteristic signature of chaotic systems, but very small, indicating mild sensitivity to initial conditions. This suggests the system resides near the edge of chaos—exhibiting diverging trajectories over time, but at a slow and predictable rate. The results align with the fractal and entropy-based findings, reinforcing a picture of structured yet dynamically rich behavior.
 
  \paragraph{Synthesis}

  Taken together, the fractal dimension, entropy measures, and LLE results paint a coherent picture of the system's dynamics. The system is neither entirely regular nor fully chaotic. The low-dimensional structure, limited entropy, and small positive LLE all point to a regime of very bounded chaos, where the system exhibits deterministic, largely predictable behavior with occasional irregularities. The observed complexity is neither trivial nor overwhelming. That is, the system is structured, yet adaptable; predictable, yet capable of surprise. These properties are characteristic of quasi-periodic attractors or bounded chaos, suggesting that the underlying mechanism, driven by the M/S ratio and bid-ask imbalance, creates a deterministic dynamics while preserving nonlinearity and feedback.


  \section{Probability of Reaching $P^*=0$}
  We know that the system has two equilibrium prices, $P^*=0$ and $P^*>0$. The key idea of the section is to train a model to predict the probability of the trajectory eventually reaching zero based on the current state, using the labels derived from the observed trajectories.
  
  \subsection{Logistic Function}
  Let $\{P_i(t)\}_{i=1}^{N}$ be a set of $N$ observed trajectories of a process $P(t)$. For each trajectory $i$, we define a binary label $y_i \in \{0, 1\}$ based on whether the trajectory reached a state considered as zero (i.e., $P(t) \le \epsilon$ for some small $\epsilon > 0$) within the observation window:
  \begin{itemize}
      \item $y_i = 0$ if trajectory $i$ reached zero.
      \item $y_i = 1$ if trajectory $i$ did not reach zero within the observation period.
  \end{itemize}
    
  For each time point $t_j$ in each trajectory $i$ ($j = 1, \dots, T_i$, where $T_i$ is the length of trajectory $i$), we create a feature vector $x_{i,j}$ representing the state at that time:
  $$x_{i,j} = \begin{pmatrix} P_i(t_j) \\ t_j \end{pmatrix}$$
  The feature vector can include the price $P_i(t_j)$ at time $t_j$, the time $t_j$ (or a normalized version), and potentially other relevant features. Our dataset consists of pairs $\{(x_{i,j}, y_i)\}$.
    
  We model the probability of reaching zero ($y=0$) given the feature vector $x$ and parameters $\theta = (w, b)$ using logistic regression:
  $$P(y=0 | x; \theta) = \frac{1}{1 + \exp(-(w^T x + b))}$$
  where:
  \begin{itemize}
      \item $x$ is the feature vector.
      \item $w$ is the weight vector associated with the features.
      \item $b$ is the bias (intercept) term.
      \item $w^T x$ is the dot product of the weight vector and the feature vector.
  \end{itemize}
    
  The parameters $\theta = (w, b)$ are estimated by minimizing the binary cross-entropy loss function over the training data:
  $$L(\theta) = - \frac{1}{M} \sum_{i=1}^{N} \sum_{j=1}^{T_i} [y_i \log(P(y=1 | x_{i,j}; \theta)) + (1 - y_i) \log(P(y=0 | x_{i,j}; \theta))]$$
  where $M = \sum_{i=1}^{N} T_i$ is the total number of data points, and $P(y=1 | x_{i,j}; \theta) = 1 - P(y=0 | x_{i,j}; \theta) = \frac{\exp(-(w^T x_{i,j} + b))}{1 + \exp(-(w^T x_{i,j} + b))}$. The optimal parameters $\hat{\theta}$ are found using an optimization algorithm.
    
  For a new state $x_{new}$ at time $t_{new}$, the estimated probability of eventually reaching zero is:
  $$\hat{P}(\text{eventually reach zero} | x_{new}; \hat{\theta}) = \frac{1}{1 + \exp(-(\hat{w}^T x_{new} + \hat{b}))}$$

  \subsection{Gradient Boosting}

Gradient Boosting Machines (GBM) for classification are an ensemble learning method that builds a strong classifier by combining the predictions of multiple weak learners, typically decision trees. The trees are built sequentially, with each new tree trying to correct the errors made by the previous ones. The "gradient" in the name refers to the fact that each new weak learner is trained on the negative gradient of the loss function with respect to the predictions of the ensemble so far.

Let the training data be denoted as $\{(x_i, y_i)\}_{i=1}^{N}$, where $x_i$ is the feature vector for the $i$-th sample, and $y_i \in \{0, 1\}$ is the binary label.

The goal is to learn a classification model $F(x)$ that predicts the probability of $y=1$ (or $y=0$).

For binary classification, a common loss function used in GBM is the negative log-likelihood of the Bernoulli distribution, which is equivalent to the logistic loss:

$$L(y, F(x)) = -[y \log(p) + (1-y) \log(1-p)]$$

where $p = P(y=1 | x) = \sigma(F(x)) = \frac{1}{1 + e^{-F(x)}}$ is the probability of $y=1$ predicted by the model, and $\sigma$ is the sigmoid function. $F(x)$ represents the raw output of the ensemble model (often called the log-odds).

The GBM builds an additive model by sequentially adding weak learners (decision trees):

$$F_M(x) = F_0(x) + \sum_{m=1}^{M} \eta h_m(x)$$

where:
\begin{itemize}
    \item $F_0(x)$ is an initial guess for the model (e.g., a constant).
    \item $M$ is the total number of trees in the ensemble.
    \item $h_m(x)$ is the $m$-th decision tree (the weak learner).
    \item $\eta$ is the learning rate ($0 < \eta \le 1$).
\end{itemize}

Initialize the model with a constant value:

$$F_0(x) = \log\left(\frac{P(y=1)}{1 - P(y=1)}\right)$$

Iteration (for $m = 1$ to $M$):

\begin{itemize}
  \item Compute Pseudo-Residuals (Negative Gradient): For each sample $i$, calculate the negative gradient of the loss function with respect to the current model $F_{m-1}(x_i)$:

  $$g_{m,i} = -\left[\frac{\partial L(y_i, F_{m-1}(x_i))}{\partial F_{m-1}(x_i)}\right]$$

  For the logistic loss, this simplifies to:

  $$g_{m,i} = y_i - p_{m-1}(x_i) = y_i - \frac{1}{1 + e^{-F_{m-1}(x_i)}}$$

  \item Fit a Weak Learner: Train a decision tree $h_m(x)$ to predict the pseudo-residuals $g_{m,i}$.
  \item Determine Leaf Values: For each leaf $j$ in the tree $h_m(x)$, determine the optimal constant value $\gamma_{m,j}$ that minimizes the loss function for the samples falling into that leaf.
  \item Update the Model: Add the new tree to the ensemble, scaled by the learning rate:

  $$F_m(x) = F_{m-1}(x) + \eta h_m(x)$$

  If $x$ falls into leaf $j$ of tree $h_m$, then $h_m(x) = \gamma_{m,j}$.
\end{itemize}

Final Prediction: After $M$ boosting rounds, the final model $F_M(x)$ is obtained. The probability of not reaching zero ($y=1$) is:

$$P(y=1 | x) = \sigma(F_M(x)) = \frac{1}{1 + e^{-F_M(x)}}$$

Then, the probability of reaching zero ($y=0$) is:

$$P(y=0 | x) = 1 - P(y=1 | x) = 1 - \frac{1}{1 + e^{-F_M(x)}} = \frac{e^{-F_M(x)}}{1 + e^{-F_M(x)}}$$

  \subsection{Calibration and Results}

  \begin{table}[h!]
    \centering
    \begin{tabular}{lcc}
        \toprule
        Metric & Logistic Regression & GBM \\
        \midrule
        Accuracy & 0.79 & 0.89 \\
        \midrule
        Confusion Matrix &
        $\begin{pmatrix}
          & \text{0} & \text{1} \\
        \text{0} & 12502 & 1697 \\
        \text{1} & 3579 & 7222
        \end{pmatrix}$
        &
        $\begin{pmatrix}
          & \text{0} & \text{1} \\
          \text{0} & 25872 & 2427 \\
        \text{1} & 3243 & 18458
        \end{pmatrix}$ \\
        \midrule
        Precision (RZ: Class 0) & 0.8805 & 0.9142 \\
        Recall (RZ: Class 0) & 0.7775 & 0.8886 \\
        Precision (NRZ: Class 1) & 0.8097 & 0.8838 \\
        Recall (NRZ: Class 1) & 0.6686 & 0.8506 \\
        \midrule
        Logistic Model's Parameters: & \multicolumn{2}{l}{$b_0=-7.2757, b_1=1.1029, b_2=5.6285$} \\ 
        \bottomrule
    \end{tabular}
    \caption{Comparison of Logistic Regression and Gradient Boosting Machines for Predicting Trajectory Reaching Zero}
    \label{tbl:prob_to_zero}
\end{table}

  Estimation results of both algorithms are shown in Table \ref{tbl:prob_to_zero}. RZ and NRZ in the table stand for "Reaching Zero" and "Not Reaching Zero". The results indicate a clear superiority of the Gradient Boosting Machines (GBM) model over Logistic Regression for predicting whether a trajectory reaches zero. The GBM achieved a significantly higher overall accuracy (0.89) compared to Logistic Regression (0.79), suggesting a more effective classification of the outcomes.

  An examination of precision and recall metrics reveals that the GBM model demonstrates better performance across both classes. For the prediction of reaching zero (Class 0), GBM exhibits higher precision (0.9142 vs. 0.8805) and notably improved recall (0.8886 vs. 0.7775), indicating a better ability to correctly identify trajectories that eventually reach zero while minimizing false positives. Similarly, for the prediction of not reaching zero (Class 1), GBM shows substantial gains in both precision (0.8838 vs. 0.8097) and recall (0.8506 vs. 0.6686), highlighting its effectiveness in correctly identifying these trajectories as well.
  
  However, unlike the GBM the parameter estimates of the Logistic Regression model offer a degree of interpretability, providing insights into the linear relationships discerned by the model between the initial feature set and the predictive output. The significantly negative intercept (-7.28) suggests a very low baseline probability of a trajectory not reaching zero under such hypothetical conditions which is likely attributable to an initial tendency for the price to decline across all observed trajectories. The positive coefficient associated with the "price" feature ($1.10$) indicates a direct relationship wherein an increase in the initial price correlates with an increased log-odds of the trajectory not reaching zero. Similarly, the positive coefficient for "normalized time" (5.63) implies that as the trajectory evolves within the observation window, the log-odds of avoiding the zero threshold also increase.
  
  The comparatively larger magnitude of the "normalized time" coefficient suggests that the progression of time has a more substantial positive impact on the likelihood of avoiding the zero threshold within the observed duration, compared to an equivalent unit increase in price. To illustrate with an example, based on the estimated parameter values of the Logistic Regression model, at time $t=700$ (which corresponds to a normalized time of $0.14$), a price around $P_{t=700}=5.88$ would result in a roughly 50\% probability of the trajectory reaching zero.\footnote{The normalized time for $t=700$ on a timeline from 1 to 5000 would be approximately $700/5000=0.14$. Requiring the price at which the model predicts a 50\% probability of reaching zero, then $-6.4877 + 1.1029 \times P=0 => P \approx 5.88$.}


  \section{A Multivariate Hidden Markov Model with 2 Hidden States}

A Hidden Markov Model (HMM) is a statistical model that describes a system assumed to be a Markov process with unobserved (hidden) states. In this case, we consider a multivariate HMM with 2 hidden states and two observed variables: Price (P) and Bid-Ask Imbalance (BA) that is defined as $BA_t = B_t - A_t$.

Let the number of hidden states be $N = 2$. Let the number of observations in a sequence be $T$. The observed sequence is $O = (o_1, o_2, ..., o_T)$, where each observation $o_t$ is a 2-dimensional vector:
$$o_t = \begin{pmatrix} P_t \\ BA_t \end{pmatrix} \in \mathbb{R}^2$$
Let the sequence of hidden states be $Q = (q_1, q_2, ..., q_T)$, where each $q_t \in \{1, 2\}$.

The HMM is defined by the following components:

The set of hidden states is $S = \{1, 2\}$.

This is a $2 \times 2$ matrix where each element $a_{ij}$ represents the probability of transitioning from hidden state $i$ at time $t$ to hidden state $j$ at time $t+1$:
$$A = \begin{pmatrix}
a_{11} & a_{12} \\
a_{21} & a_{22}
\end{pmatrix}$$
where $a_{ij} = P(q_{t+1} = j | q_t = i)$, and for each row $i \in \{1, 2\}$:
\begin{itemize}
    \item $0 \leq a_{ij} \leq 1$ for all $j \in \{1, 2\}$
    \item $\sum_{j=1}^{2} a_{ij} = 1$
\end{itemize}

For a multivariate HMM with Gaussian emissions, for each state $j \in \{1, 2\}$, we have a multivariate Gaussian distribution with a mean vector $\boldsymbol{\mu}_j$ and a covariance matrix $\boldsymbol{\Sigma}_j$. For our two observed variables, the mean vector for state $j$ is:
$$\boldsymbol{\mu}_j = \begin{pmatrix} \mu_{P,j} \\ \mu_{BA,j} \end{pmatrix}$$
and the covariance matrix for state $j$ is a $2 \times 2$ matrix. Given the provided results, the covariance matrices are diagonal, indicating that Price and Bid-Ask Imbalance are assumed to be uncorrelated within each state:
$$\boldsymbol{\Sigma}_j = \begin{pmatrix} \sigma_{P,j}^2 & 0 \\ 0 & \sigma_{BA,j}^2 \end{pmatrix}$$
The probability density function (PDF) of observing $o_t = \begin{pmatrix} P_t \\ BA_t \end{pmatrix}$ given that the hidden state is $j$ is:
$$b_j(o_t) = P(o_t | q_t = j) = \frac{1}{(2\pi)^{2/2} |\boldsymbol{\Sigma}_j|^{1/2}} \exp\left(-\frac{1}{2} (o_t - \boldsymbol{\mu}_j)^T \boldsymbol{\Sigma}_j^{-1} (o_t - \boldsymbol{\mu}_j)\right)$$
For a diagonal covariance matrix, this simplifies to the product of two univariate Gaussian PDFs:
$$b_j(o_t) = \left( \frac{1}{\sqrt{2\pi\sigma_{P,j}^2}} \exp\left(-\frac{(P_t - \mu_{P,j})^2}{2\sigma_{P,j}^2}\right) \right) \times \left( \frac{1}{\sqrt{2\pi\sigma_{BA,j}^2}} \exp\left(-\frac{(BA_t - \mu_{BA,j})^2}{2\sigma_{BA,j}^2}\right) \right)$$
The set of emission parameters is $B = \{(\boldsymbol{\mu}_1, \boldsymbol{\Sigma}_1), (\boldsymbol{\mu}_2, \boldsymbol{\Sigma}_2)\}$.

This is a $1 \times 2$ vector that defines the probability of being in each hidden state at the beginning of the sequence (time $t=1$):
$$\pi = (\pi_1, \pi_2)$$
where $\pi_i = P(q_1 = i)$, and:
\begin{itemize}
    \item $0 \leq \pi_i \leq 1$ for all $i \in \{1, 2\}$
    \item $\sum_{i=1}^{2} \pi_i = 1$
\end{itemize}

The multivariate HMM with 2 hidden states and observations of Price (P) and Bid-Ask Imbalance (BA) is defined by the parameter set $\lambda = (A, B, \pi)$.

\subsection{Estimation Results and Comments}

\begin{table}[h!]
    \centering
    \begin{tabular}{l l l c}
        \toprule
        \textbf{Parameter} & \textbf{State} & \textbf{Transition} & {\textbf{Value}} \\
        \midrule
        \multirow{4}{*}{\textbf{Transition Probabilities (A)}}
            & \multirow{2}{*}{From State 1} & To State 1 & 0.99720110 \\
            &                                         & To State 2 & 0.00279890 \\
            \cmidrule{2-4}
            & \multirow{2}{*}{From State 2} & To State 1 & 0.00092425 \\
            &                                         & To State 2 & 0.99907575 \\
        \midrule
        \multirow{4}{*}{\textbf{Means (B)}}
            & \multirow{2}{*}{State 1} & Price (P)   & -9.59565388e-123 \\ 
            &                                     & Bid - Ask  & 0.00000000e+0 \\ 
            \cmidrule{2-4}
            & \multirow{2}{*}{State 2} & Price (P)   & -0.00188545 \\
            &                                     & Bid - Ask  & 120.139187 \\
        \midrule
        \multirow{4}{*}{\textbf{Variances (B)}}
            & \multirow{2}{*}{State 1} & Price (P)   & 1.83339622e-7 \\ 
            &                                     & Bid - Ask  & 1.83339622e-7 \\ 
            \cmidrule{2-4}
            & \multirow{2}{*}{State 2} & Price (P)   & 0.00006810 \\
            &                                     & Bid - Ask  & 13046.9587 \\
        \midrule
        \multirow{2}{*}{\textbf{Initial Probabilities ($\pi$)}}
            & State 1                    &             & 0.0 \\
            & State 2                    &             & 1.0 \\
        \bottomrule
    \end{tabular}
    \caption{Estimated HMM Parameters}
    \label{tbl:estimated_hmm}
\end{table}

Results of the multivariate Hidden Markov Model with two hidden states are shown in Table \ref{tbl:estimated_hmm}. Considering both price change and bid-ask imbalance, identifies two highly persistent regimes.

\textbf{State 1:} Characterized by a mean price change very close to zero and a mean bid-ask imbalance of zero, this state exhibits very low variances for both variables ($0.000000183$). The transition probabilities show extremely high persistence ($0.997$) with a very low probability of transitioning to State 2 ($0.00279$). This state strongly suggests the equilibrium at $P^*=0$, indicating a stable condition where both price changes and the forces reflected in bid-ask imbalance are minimal.

\textbf{State 2:} This state displays a small negative mean price change ($-0.001885$) and a high positive mean bid-ask imbalance ($120.14$). The variances are higher compared to State 1, with $0.0000681$ for price change and $13000$ for bid-ask imbalance. Similar to State 1, this state also shows very high persistence ($0.999$) with a very low probability of transitioning to State 1 ($0.000924$). This state likely represents a regime where the price is above zero ($P^*>0$), experiencing a slight downward drift while there is a significant buying pressure indicated by the positive bid-ask imbalance. The high persistence suggests that the system can remain in this state for extended periods.

The model indicates that all trajectories start in State 2 ($1.0$), the state associated with a positive price level and buying imbalance.

State 1 strongly corresponds to the equilibrium at $P^*=0$ due to its near-zero means and very low variances for both price change and bid-ask imbalance, along with its high persistence. This suggests a stable and absorbing-like state at the zero price level.

State 2 appears to represent a persistent state associated with the $P^*>0$ regime. The negative mean price change suggests a tendency towards the zero equilibrium, while the high positive bid-ask imbalance indicates ongoing buying pressure. The very high persistence of this state implies that the system can remain in this positive-price regime for a long time before potentially transitioning to the zero-price equilibrium.

The extremely low transition probabilities between the two states suggest that the system tends to stay within one of these regimes for extended durations. This might imply that the transition from a positive price level with buying imbalance to the zero-price equilibrium is a rare event within the observed time frame or under the modeled dynamics.

The initial condition of starting solely in State 2 indicates that the observed trajectories begin in the positive-price regime, which is consistent with the simulation procedure, given that all started from the same initial conditions.

Overall, this two-state model provides a simplified but clear distinction between a stable zero-price equilibrium and a persistent state characterized by a positive price level with buying imbalance and a gradual downward drift. The high persistence within each state and the low transition probabilities suggest that the system tends to reside in one of these conditions for considerable periods.


 \section{Conclusion}
 Paper studies a bifurcative nature of price trajectories that are a result of trading activities of endogenous and autonomous agents who trade by placing bid and ask orders to the limit order book (LOB) within an artificial stock market exchange (ASME). Modeling trading activities at a microstructure level within an agent-based framework is not novel approach and the current paper is an extension of our prior work. Therein it was shown that trading activities of myopic, random agents, like those within the current paper, push the price of an asset into the direction of the money-to-stock ratio that works as a drift, with a market bid-ask imbalance affecting the "noise". 

 The current study extends this theoretical foundation by systematically exploring the bifurcation structure inherent to the system. Specifically, we show that the ASME exhibits bistability, whereby qualitatively distinct long-run equilibria, a degenerate state at $P^*=0$ and a persistent positive price equilibrium $P^*>0$, can emerge from identical initial conditions. This divergence of trajectories into distinct basins of attraction underscores the nonlinear and path-dependent nature of price evolution in the model. Between the two attractor basins lies a metastable region, functioning as a transient phase space, where trajectories exhibit heightened sensitivity to stochastic perturbations.

 The period around the bifurcation point is characterized by elevated volatility, suggesting that the market is more sensitive to endogenous fluctuations. The trajectories converging toward $P^*=0$ exhibit protracted transitions and greater variance, suggesting that the path to the terminal, absorbing state is dynamically extended. In contrast, trajectories associated with the $P^*>0$ regime display ongoing oscillatory behavior around a stochastic price level, lacking convergence to a fixed price. This behavior is in line with our previous findings, where price trajectories of the high-price regime were shown to fluctuate within a narrow band.

 Values of Lyapunov exponent are consistently positive, which is a characteristic signature of chaotic systems, but very small which does not make a conclusive proof for the chaos. The system appears to exhibit an intermittent chaotic behavior or low-dimensional chaos, where the market exhibits complexity and irregularity, before eventually settling into a non-chaotic state. The view aligns with the fractal and entropy-based findings, reinforcing a picture of structured yet dynamically rich behavior.

 The consistent fractal dimension also points to potential predictability of certain aspects of the system, as the dynamics appear to be governed by a set of low-dimensional attractors. However, the nonlinearity and apparent irregularity suggest that complete prediction remains challenging, with the system exhibiting deterministic, largely predictable behavior with occasional irregularities.
 
 In sum, the study contributes a novel perspective on the inherent path dependence and complex dynamics of artificial stock markets. It highlights the importance of micro-level interactions in shaping macro-level market outcomes. The results of the paper provide compelling evidence that the ASME dynamics, governed by minimalistic, zero-intelligent agent rules and microstructural constraints, naturally give rise to nonlinear bifurcations, multistability, and path dependence. 

 \printbibliography

\section*{Disclosure of interest}
The authors declare that they have no conflict of interests relevant to this publication.

\section*{Funding}
No funding was received for this study.


\section*{Appendix}
\begin{appendices}

\section{The Separatrix Estimation Algorithm: Classification Error Minimization}
\label{app:separatrix_classification}

Let \(X = \{ x_i(t) \}\) be a set of one-dimensional trajectories, where each trajectory \(x_i(t)\) evolves over time \(t\) according to an underlying dynamical system.

For each trajectory \(x_i(t)\), we define its final state as:
$$
S_i = \lim_{t \to T} x_i(t),
$$
where \(T\) is the final observation time. In practice, we approximate this limit by computing the mean over the last \(L\) time steps:
$$
\bar{x}_i = \frac{1}{L} \sum_{k=T-L}^{T} x_i(k).
$$

We classify each trajectory into three categories:
\begin{itemize}
    \item \(S_i = 0\) if \(|\bar{x}_i| < \epsilon_0\) (converging to zero threshold)
    \item \(S_i = 1\) if \(\bar{x}_i > \epsilon_1\) (large positive threshold)
    \item \(S_i = -1\) otherwise (undetermined)
\end{itemize}
where \(\epsilon_0\) and \(\epsilon_1\) are predefined thresholds.

Define sets:
$$
Z = \{ i : S_i = 0 \}, \quad P = \{ i : S_i = 1 \}.
$$

If either \(Z\) or \(P\) is empty, no separatrix is found.

We define a time index \(t_c\) that is approximately halfway through the trajectories:
$$
t_c = \frac{1}{2} \text{median} \{ T_i : i \in Z \cup P \}.
$$
where \(T_i\) is the length of trajectory \(x_i(t)\).

For each \(i \in Z \cup P\), we record \(x_i(t_c)\), forming two groups:
$$
X_Z = \{ x_i(t_c) : i \in Z \}, \quad X_P = \{ x_i(t_c) : i \in P \}.
$$

Define the sorted set of unique values from \(X_Z \cup X_P\):
$$
X_s = \{ x_{(1)}, x_{(2)}, \dots, x_{(m)} \}, \quad x_{(1)} < x_{(2)} < \dots < x_{(m)}.
$$

We consider candidate separatrix points as midpoints between consecutive values:
$$
s_j = \frac{x_{(j)} + x_{(j+1)}}{2}, \quad j = 1, \dots, m-1.
$$

For each candidate \(s_j\), we define groups:
$$
G_1 = \{ x_i \leq s_j \}, \quad G_2 = \{ x_i > s_j \}.
$$

Define classification errors:
$$
E_j = \frac{| \{ i \in G_1 : i \in P \} | + | \{ i \in G_2 : i \in Z \} |}{|Z| + |P|}.
$$

The optimal separatrix \(s^*\) minimizes \(E_j\):
$$
s^* = \arg\min_{j} E_j.
$$

The method estimates a separatrix by identifying a threshold that best separates trajectories based on their final states, minimizing classification errors.


\section{The Separatrix Estimation Algorithm: Entropy-Based Estimation}
\label{app:separatrix_entropy}

Let \( \mathcal{X} = \{ P_i(t) \}_{i=1}^N \) be a set of one-dimensional trajectories, where each \( P_i(t) \) evolves over discrete time \( t \in \{0, 1, \dots, T-1\} \) according to an underlying dynamical system.

Given a total number of time points \( T \), we select an evaluation time index \( t_c \) based on a fraction \( f \in [0, 1] \) of the total duration:
\[
t_c = \lfloor f (T-1) \rfloor.
\]
If \( t_c \geq T \), we set \( t_c = T-1 \).

The state values at this time are given by the set:
\[
\mathcal{P}_c = \{ P_i(t_c) \mid i = 1, \dots, N \},
\]
where \( N \) is the number of trajectories. We consider only the finite and non-NaN values in this set.

We define the minimum and maximum state values at \( t_c \) among the non-NaN values:
\[
P_{\min} = \min \{ p \in \mathcal{P}_c \}, \quad P_{\max} = \max \{ p \in \mathcal{P}_c \}.
\]
We partition the interval \( [P_{\min}, P_{\max}] \) into \( B = \texttt{num\_bins} \) bins of equal width \( \Delta P = \frac{P_{\max} - P_{\min}}{B} \). The boundaries of the bins are given by:
\[
b_k = P_{\min} + k \Delta P, \quad k = 0, 1, \dots, B.
\]
The \( k \)-th bin is the interval \( [b_k, b_{k+1}) \), for \( k = 0, 1, \dots, B-1 \), with the last bin being \( [b_{B-1}, b_B] \).

Define the histogram \( H_k \) as the number of trajectories whose state value at time \( t_c \) falls into the \( k \)-th bin:
\[
H_k = \sum_{i=1}^N \mathbb{I}(b_k \leq P_i(t_c) < b_{k+1}), \quad \text{for} \quad k = 0, 1, \dots, B-1,
\]
where \( \mathbb{I}(A) \) is the indicator function, equal to 1 if the condition \( A \) is true, and 0 otherwise. For the last bin \( k = B-1 \), the condition is \( b_{B-1} \leq P_i(t_c) \leq b_B \).

The empirical probability mass function \( p_k \) for each bin is estimated by normalizing the histogram by the total number of trajectories \( N_{\text{valid}} \) that have a non-NaN value at time \( t_c \):
\[
p_k = \frac{H_k}{N_{\text{valid}}}, \quad \text{for} \quad k = 0, 1, \dots, B-1,
\]
where \( N_{\text{valid}} \) is the number of elements in \( \mathcal{P}_c \).

The Shannon entropy \( S \) of the discrete probability distribution \( \{p_k\}_{k=0}^{B-1} \) is calculated using the base-2 logarithm:
\[
S = - \sum_{k=0}^{B-1} p_k \log_2 p_k.
\]
By convention, we define \( 0 \log_2 0 = 0 \) to handle cases where a bin has zero probability. The entropy \( S \) quantifies the uncertainty or spread of the distribution of trajectory states at the evaluation time \( t_c \).

The code estimates the separatrix \( s^* \) as the median of the state values of the trajectories at the evaluation time \( t_c \):
\[
s^* = \text{median}(\mathcal{P}_c).
\]
The rationale behind using the median is that at a time when the trajectories are most spread out due to the influence of the separatrix, the median provides a measure of the central tendency of this distribution, which is heuristically assumed to be related to the location of the separatrix.

This method estimates the separatrix by analyzing the spatial distribution of an ensemble of trajectories at a specific time \( t_c \). The Shannon entropy is used to quantify the spread of this distribution, and the median of the trajectory values at that time serves as a point estimate for the separatrix. The choice of the evaluation time \( t_c \) and the number of bins \( B \) can influence the result. The method relies on the assumption that the separatrix is located in a region of the state space where the trajectories exhibit maximum spread at some intermediate time.


\section{The Separatrix Estimation Algorithm: SVM-Based Estimation}
\label{app:separatrix_svm}

Let
\[
\{x_i(t)\}_{i=1}^N \quad (x_i(t) \in \mathbb{R})
\]
be a collection of trajectories observed over time \( t \). We assume that for each trajectory \( x_i(t) \), there exists a time interval near the end of the observation period (last \( L \) time points) where the trajectory converges. That is, for a trajectory \( x_i(t) \) with \( t \in [0, T] \), define the \emph{converged value} as
\[
\bar{x}_i = \frac{1}{L} \sum_{k=T-L+1}^{T} x_i(t_k).
\]

We label trajectories according to their converged values:
\begin{enumerate}
    \item If
    \[
    \left| \bar{x}_i \right| < \epsilon_0,
    \]
    then assign label \( l_i = -1 \) (i.e., the trajectory converges to near zero).
    \item If
    \[
    \bar{x}_i > \epsilon_1,
    \]
    then assign label \( l_i = 1 \) (i.e., the trajectory converges to a large state).
    \item Otherwise, the trajectory is considered \emph{undetermined} and is excluded from further analysis.
\end{enumerate}

Let the set of indices for the labeled trajectories be denoted by
\[
I = \{ i \mid l_i \in \{-1, 1\} \}.
\]

At a designated comparison time \( t_c \) (or index \( t_c \)), we record the state of each labeled trajectory:
\[
p_i = x_i(t_c), \quad i \in I.
\]
These values \( \{p_i\} \) are used to determine the separatrix.

We seek a separatrix value \( s \in \mathbb{R} \) that ideally divides the state space such that:
\begin{itemize}
    \item For trajectories with \( l_i = -1 \), we desire \( p_i < s \).
    \item For trajectories with \( l_i = 1 \), we desire \( p_i > s \).
\end{itemize}
To enforce a margin \( \delta > 0 \), we define the individual cost for each trajectory as follows:
\begin{align*}
    C_i(s) &= \max \{0, \, p_i - s + \delta \} \quad \text{if } l_i = -1, \\
    C_i(s) &= \max \{0, \, s - p_i + \delta \} \quad \text{if } l_i = 1.
\end{align*}
The total cost function is then given by
\[
C(s) = \sum_{i \in I} C_i(s) = \sum_{\substack{i \in I \\ l_i = -1}} \max \{0, \, p_i - s + \delta\} + \sum_{\substack{i \in I \\ l_i = 1}} \max \{0, \, s - p_i + \delta\}.
\]

The goal is to find the separatrix \( s^* \) that minimizes the cost function:
\[
s^* = \arg\min_{s \in [s_{\min}, s_{\max}]} C(s),
\]
where the search bounds \( s_{\min} \) and \( s_{\max} \) are chosen based on the observed range of \( \{p_i\} \) (typically with some extension to ensure the optimum is captured).


\section{Fractal Dimension: Mathematical Description}

The fractal dimension provides a quantitative measure of the complexity of a set or time series, reflecting how it fills the space it occupies. For time series data, the correlation dimension is commonly used, which is defined through the correlation sum.

Given a time series \( X(t) \) and a corresponding m-dimensional embedding, the steps to calculate the correlation dimension are as follows:

The time series \( X(t) \) is embedded into an m-dimensional space using a time delay \( \tau \). Each vector in this phase space is given by:
\[
\mathbf{x}_i = \left( X(i\tau), X((i+1)\tau), \dots, X((i+m-1)\tau) \right)
\]
where \( i = 0, 1, \dots, N-m \), and \( N \) is the total number of data points in the series.

The distance between any two vectors in the phase space is calculated, typically using the Euclidean distance:
\[
d(\mathbf{x}_i, \mathbf{x}_j) = \sqrt{\sum_{k=0}^{m-1} \left(X(i+k\tau) - X(j+k\tau)\right)^2}
\]
for \( i \neq j \).

The correlation sum \( C(r) \) is computed as the proportion of pairs of points whose distance is less than a threshold \( r \):
\[
C(r) = \frac{1}{N(N-1)} \sum_{i \neq j} \Theta(r - d(\mathbf{x}_i, \mathbf{x}_j))
\]
where \( \Theta \) is the Heaviside step function, which equals 1 if the distance is smaller than \( r \), and 0 otherwise.

The correlation dimension is determined by the scaling behavior of the correlation sum \( C(r) \) as \( r \) approaches 0:
\[
C(r) \sim r^\alpha
\]
where \( \alpha \) is the correlation dimension. The fractal dimension is given by:
\[
D_2 = \lim_{r \to 0} \frac{\ln C(r)}{\ln r}
\]
This value \( D_2 \) describes how the number of points in the phase space scales with the distance threshold \( r \), providing insight into the underlying fractality of the system.

\end{appendices}

\end{document}